# ONTOLOGY BASED DOCUMENT CLUSTERING USING MAPREDUCE

Abdelrahman Elsayed [1], Hoda M. O. Mokhtar [2] and Osama Ismail[3]

[1]Central Lab for Agriculture Expert Systems, Agriculture Research Center, Egypt
[2,3]Faculty of Computers and Information, Cairo University, Egypt

## ABSTRACT

*Nowadays, document clustering is considered as a data intensive task due to the dramatic, fast increase in the number of available documents. Nevertheless, the features that represent those documents are also too large. The most common method for representing documents is the vector space model, which represents document features as a bag of words and does not represent semantic relations between words. In this paper we introduce a distributed implementation for the bisecting k-means using MapReduce programming model. The aim behind our proposed implementation is to solve the problem of clustering intensive data documents. In addition, we propose integrating the WordNet ontology with bisecting k-means in order to utilize the semantic relations between words to enhance document clustering results. Our presented experimental results show that using lexical categories for nouns only enhances internal evaluation measures of document clustering; and decreases the documents features from thousands to tens features. Our experiments were conducted using Amazon Elastic MapReduce to deploy the Bisecting k-means algorithm.*

## 1. INTRODUCTION

Document clustering is the process of grouping similar documents. Document clustering helps in organizing documents and enhances the methods of displaying search engine results. "Grouper" is an example of document clustering interface, it groups the results of search dynamically into clusters, and it can be integrated with any search engine [1]. In most document clustering approaches, documents are represented using vector space model [2]. Given a set of *n* documents {d1, d2, d3,...., dn}, the unique terms *m* are extracted from these documents, and the documents are represented as $n \times m$ Matrix. Each row of the matrix represents a document vector, and columns represent document terms. In vector space model, terms are represented as a bag of words. The main drawback of this representation is that it ignores the semantic relations between terms [3].

Ontology has been adopted to enhance document clustering, further it has been used to solve the problem of large number of document features [4]. Inspired by the importance of ontology and its capability to enhance document clustering, in this research we investigate the use of WordNet ontology [5] in order to decrease the huge volume of document features to just 26 features representing the WordNet lexical noun categories. Moreover, WordNet helps in representing semantic relations between terms. For example, the words **cheese** and **bread** have the same lexical noun category ***noun.food***, so it will be represented by the same term feature, which consequently results in dramatically reducing the number of dimensions.

Although using WordNet to reduce document dimensionality enhances document clustering efficiency, the traditional implementation of this approach is not very efficient due to the huge volumes of documents. Thus, using a parallel programming paradigm for implementing this solution turns to be a more appealing approach. Thus, we investigated running document clustering process in a distributed framework, by adopting the common MapReduce programming model [6]. MapReduce is a programming model initiated by Google's Team for processing huge datasets in distributed systems. It presents simple and powerful interface that enables automatic parallelization and distribution of large-scale computations. In addition, it enables programmers who have no experience with parallel and distributed system to utilize the resources of a large distributed system in easily and efficient way [6].

Also, we integrate WordNet ontology with bisecting k-means algorithm using MapReduce in document clustering.

The reminder of this paper is structured as follows: Section 2 introduces related work and a brief background about K-means and its variant the bisecting k-means. Section 3 introduces the problem and discusses the proposed system. Section 4 is devoted to explain the document clustering evaluation methods that we used to evaluate our proposed clustering approach. Section 5 introduces the details about the used datasets and discusses the experimental results. Finally, Section 6 concludes and presents possible directions for future work.

## 2. BACKGROUND AND RELATED WORK

Ontology is usually referred to as a method for obtaining an agreement on similar activities in a specific domain. [7]. Many ontologies have been developed for different domains, such as food ontology [8], gene ontology [9], and agriculture ontology [10]. In the following discussion we overview some research efforts that investigated the use of ontology for enhancing the process of documents clustering.

In [11], the authors integrated WordNet ontology with document clustering. WordNet organizes words into sets of synonyms called 'synsets'. They tried to solve the problem of the bag of words representation, in order to represent relationships between terms which do not co-occur literally. The authors considered the synsets as concepts and extended the bag of words model by including the parent concepts (hypernyms) of synsets up to five levels. For example, they utilized the WordNet ontology to find the similarity between the related terms, such as "beef" and "pork", these two terms have the same parent concept "meat". Therefore, a document having the term "beef" will be related to a document with the term "pork" appearing in it; the proposed approach is shown to enhance the clustering performance.

In [12], the authors extended Hotho experiment by adding all word senses instead of the first sense; also they included all hypernym levels instead of just 5 levels of hypernym. They concluded that, their approach does not enhance document clustering, because more noises are attached to the dataset by including all word senses and all hypernym levels. In [13], the authors claim that the drawback of augmenting WordNet synsets with original terms is increasing the dimensionality of terms. In [13] they used ontology to reduce the number of features. In their work they showed that by identifying and using noun features alone text document clustering is improved. Furthermore, they used word sense disambiguation techniques in order to resolve polysemy, which happens when the word has different meanings. In addition, they used ontology to resolve synonymy problems, which occurs when two words refer to the same concept, by using the corresponding concepts and expelling the synonymous.

Recupero [4] coped with problems of vector space model, which include high dimensionality of data and ignoring the semantic relationships between terms. Recupero used WordNet ontology to find the lexical category of each term, and replaced each term with its corresponding lexical category. Using the fact that WordNet has 41 lexical categories for representing nouns and verbs, all documents dimensions were reduced to just 41 dimensions. Recupero used ANNIE, which is an information extraction system [14] to help in understanding if two words or compound words refer to the same entity. For example entities like "Tony Blair, Mr. Blair, Prime Minister Tony Blair, and Prime Minister" will all be represented by one entity.

In the remaining of this section we present previous work in document clustering techniques. K-means clustering and its variant bisection k-means are the focus of our discussion due to their wide use and adoption in many applications.

## 2.1 K-means and Bisecting k-means

K-means is a widely known partitioning clustering technique, The input of k-means is N objects, M dimensions, and k representing the targeted number of clusters. The output of k-means is k clusters represented by the cluster centers. The main idea in the traditional k-means algorithm is to increase (maximize) inter cluster similarity and minimize the similarity with other clusters [15] [16].

K-means is an iterative algorithm. The steps of k-mean are repeated until convergence, which may be satisfied if there is no changes in cluster centers (more specifically when convergence is achieved), or may be satisfied after a specific number of iterations.

Bisecting k-means is a variant of k-means developed by Steinbach, et al [17]; it has been applied in document clustering. In [17] the authors showed that bisecting k-means gives better document clustering results than the basic k-means algorithm. In brief, bisecting k-means proceeds as follows: Given a desired number of clusters denoted by "k":

1- Use basic k-means algorithm to get two sub clusters of input dataset.
2- Select the sub-cluster that satisfies the highest overall similarity to be the input dataset.
3- Repeat steps 1, 2 until reaching the desired number of clusters.

Choosing which cluster to be split, may be done by either selecting the largest cluster or selecting the cluster that has the least overall similarity. In [17] the authors concluded that the difference between the two methods is small.

## 3. PROBLEM STATEMENT AND PROPOSED SOLUTION

The two main problems which face document clustering process are namely: the huge volume of documents, and the large size of document features. In this paper we propose a novel approach to enhance the efficiency of document clustering. The proposed system is divided into two phases; the objective of the first phase is to decrease the very large number of document terms (also known as document features) through adopting WordNet ontology. The second phase aims to cope with the huge volume of documents by employing the MapReduce parallel programming paradigm to enhance the system performance. In the following discussion we discuss each phase in more details.

## 3.1 Phase 1: Using WordNet Ontology for Document Features Reduction

Ontology plays a vital role in document clustering process by decreasing the large number of documents features from thousands to tens of features only. The features reduction process utilizes ontology characteristic which includes semantic relations between words such as synonyms and hierarchical relations between words. From hierarchy relations we can get word parent and use it for representing document features. For example the words corn, wheat, and rice can be represented by only one word which is plant. Also words such as beef and pork can be represented by words meat or food depending of the degree of hierarchy that will be used in the clustering process. Exploiting semantic relations between words will help in setting documents that contain words such as rice, wheat and corn at the same cluster. This paper utilizes semantic relations that are included in WordNet ontology as follow:

WordNet organizes word synsets into 45 lexicographer files; which are further divided into 26 for nouns, 15 for verbs, and 4 for adverbs and adjectives. We extended [4] approach which replaces each term by its corresponding WordNet lexical category, we get the lexical category which corresponds to nouns terms only. Also we used lexical frequency inverse document frequency "lf-idf" instead of lexical frequency to give low weight for lexical categories that occur in most of the documents. Figure (1) displays the detailed steps of replacing the traditional representation of document terms as a bag of words by a bag of lexical categories.

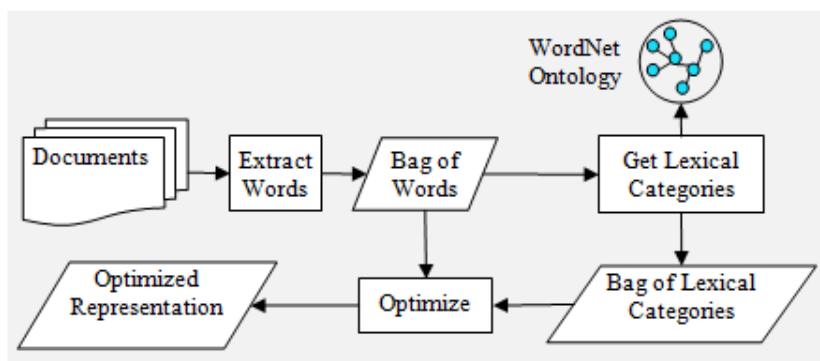

Figure 1 Steps for representing documents as bag of lexical categories

Given a set of documents, the first step in this phase is the Extract-Words process. The Extract-Words process removes stop words and extracts the remaining words, it generates two files; the vocabulary file that contains the list of all words; and the document-words file which stores associations between words and document (bag of words). Figure (2) displays the format of the resulting document-words file [18].

| | |
|---|---|
| **D** | //Number of documents |
| **W** | //Number of words in the vocabulary |
| **NNZ** | // Number of nonzero counts in the bag-of-words |
| **docID wordID count** | //Document identifier , word identifier , and count |
| **docID wordID count** | |
| **docID wordID count** | |
| **docID wordID count** | |

Figure 2 Format of bag of words file

The next process is "Get Lexical Categories"; it converts the bag of words into a bag of WordNet lexical categories. This process involves the following steps:

1- Mapping each word to its WordNet lexical category. This step generates a WordID-CategoryID file. In case of the word doesn't have a corresponding WordNet lexical category, it is mapped to Uncategorized category.

2- Generating a bag of lexical categories that replaces " docID wordID count " to "docID CategoryID count".

The third process is "Optimize", the input to this process is the bag of words file or bag of lexical categories, and the output is an optimized representation. In case of bag of words, each document will be represented by one line as follows:

 "docID word1ID:count word2ID:count ...... wordnID:count ".

In case of bag of lexical categories, each document will be represented also by one line as follows:

"docID Lexical1ID:count Lexical2ID:count ...... LexicalnID:count ".

The optimized representation of bag of word reduces file size dramatically. For example, for "PubMed" bag of words dataset [18]: it reduced the bag of words file from 6.3 gigabytes to 928.475 megabytes.

### 3.2 Phase 2: Bisecting k-means Implementation over MapReduce Framework

To overcome the continuous increase in document size, we used MapReduce to run the bisecting k-means algorithm. To implement bisecting k-means over the MapReduce framework, first, traditional k-means algorithm is implemented to generate two clusters; we adapt the method which is presented in [19] as follow:

1. Initialize centers.

2. In Map function each document vector is assigned to the nearest center. The key of map function is the document Id and the value is document vector, the map function emits cluster index, and associated document vector.

3. In Reduce function new centers are calculated. The key will be the cluster index and the value is document vector. Using cluster index instead of Cluster centroid in reduce function as a key reduce the amount of data that will be aggregated by reduce workers nodes.

4. In the clear function of reducer, new centers are saved in a global file. It will be used for next iteration of k-means algorithm.

5. Finally, if convergence is achieved then stop, else go to step 2.

The main idea of implementing the bisecting k-means algorithm to run on MapReduce is controlling Hadoop Distributed File System file paths of dataset, cluster centers, and output clusters. Controlling HDFS paths help in implementing algorithms that use multiple MapReduce iterations such as Bisecting k-means algorithm.

Figure (3) displays the algorithm of running bisecting k-means on MapReduce. The inputs of the algorithm are as follows: the path of input dataset, the path of cluster centroids, the output path, number of dimensions, and the required number of clusters. The output of the algorithm is the documents' clusters, and clusters centroids.

The proposed algorithm proceeds as follows: First, it calls the basic k-means with the following parameters: path of dataset, path of cluster centroids, path of output, and number of features (dimensions). In this step, the basic k-means algorithm is adjusted to generate two clusters at each call. Next, the largest cluster is selected to be the path of the input dataset. Also, the string that represents the output path is concatenated with "1", and the string that represents the path of clusters' centroids is concatenated with "1". Every time we call the basic k-means the number of obtained clusters is increased by one except at the last call. The process of calling the basic k-means algorithm is repeated until we reach the desired number of clusters.

The following figure shows the pseudo-code of the proposed algorithm, Let $P_I$ represents the path of the input dataset, k denotes the desired number of clusters, $N_d$ denotes the number of dimensions, $P_{CC}$ represents the path of the clusters' centroids, and $P_O$ represents the output path. Let Basic-K-means be the procedure that computes the traditional k-means algorithm and return k-clusters.

```
Bisecting k-means(PI,k,Nd,PCC,PO)
Output: Clusters' centers,  set of clusters.
      BEGIN:
                  R= 0 // R is number of clusters obtained
                  FirstTime = true //the first time of calling basic
                                    k-means
             while (R < K)
                BEGIN
                     Basic-K-means(PI,Nd,PCC,PO);
                     if (FirstTime)
                              R = R+ 2;
                              FirstTime = false;
                      else
                              R = R+1;

                     PI = Path of Largest cluser returned by
                              Basic-K-means
                     PCC= concatenate(PCC,"1");
                     PO = concatenate(PO,"1");
                END
      END
```

Figure 3 Bisecting k-means over MapReduce Algorithm

# 4. DOCUMENT CLUSTERING EVALUATION METHODS

For evaluating the proposed document clustering approach, we perform two types of cluster evaluation; external evaluation, and internal evaluation. External evaluation is applied when the documents are labeled (i.e. their true cluster are known apriori). Internal evaluation is applied when documents labels are unknown.

## 4.1 External Evaluation

Three measures for external evaluation of clustering results are used in this paper. These measures are purity, entropy, and the F-measure [20], [21], [17]. As the value of purity and F-measure increase it means that better clustering is achieved,  on the other hand,  as the value of entropy decreases it means better results are achieved [22]. In the rest of this section we discuss each measure in more details.

## Purity

Purity is a measure for the degree at which each cluster contains single class label [20]. To compute purity, for each cluster $j$, we compute the number of occurrences for each class $i$ and select the maximum occurrence $(\mathtt{Max}_{ij})$, the purity is thus the summation of all maximum occurrences $(\mathtt{Max}_{ij})$ divided by the total number of objects $n$.

$$P = \frac{1}{n} \sum_{j}^{c} \text{Max}_{ij}$$

## Entropy

Entropy is a measure of uncertainty for evaluating clustering results [21]. For each cluster $j$ the entropy is calculated as follow

$$E(j) = \sum_{i=1}^{c} P_{ij} \log_2 \frac{1}{P_{ij}}$$

Where, $c$ is the number of classes, $P_{ij}$ is the probability that member of cluster $j$ belongs to class $i$, $P_{ij} = \frac{n_{ij}}{n_j}$, where $n_{ij}$ is the number of objects of class $i$ belonging to cluster $j$, $n_j$ is total number of objects in cluster $j$.

The total entropy $E$ for all clusters is calculated as follow,

$$E = \sum_{j=1}^{k} \frac{n_j E(j)}{n}$$

Where $k$ is the number of clusters, $n_j$ is the total number of objects in cluster $j$, and $n$ is the total number of all objects.

## F-measure

F-measure is a measure for evaluating the quality for hierarchical clustering [4]. F-measure is a mix of recall and precision. In [17] they considered each cluster as a result of a query, and each class is considered as the desired set of documents. First the precision and recall are computed for each class $i$ in each cluster $j$.

$$Recall(i,j) = \frac{n_{ij}}{n_i}$$

$$Precision(i,j) = \frac{n_{ij}}{n_j}$$

Where, $n_{ij}$ is the number of objects of class $i$ in cluster $j$, $n_i$ is total number of objects in class $i$ and $n_j$ is the total number of objects in cluster $j$.

The F-measure of class $i$ and cluster $j$ is then computed as follow

$$F(i,j) = \frac{(2 * Recall(i,j) * Precision(i,j))}{Recall(i,j) + Precision(i,j)}$$

the maximum value of F-measure of each class is selected then, the total f-measure is calculated as following, where $n$ is total number of documents, $c$ is the total number of classes

$$F = \sum_{i=0}^{c} \frac{n_i}{n} \, \text{Max} \, F(i,j)$$

## 4.2 Internal evaluation

For internal evaluation, the goal is to maximize the cosine similarity between each document and its associated center [23]. Then, we divide the results by the total number of documents.

$$\text{Maximize} \frac{\sum_{j=1}^{k} \sum_{d=0}^{n_j} \cos(d_j, d)}{n}$$

Where $k$ denotes to number of clusters, $n_j$ is the number of documents assigned to cluster $j$, $d_j$ is the center of cluster $j$, $d$ is the document vector. The value of this measure is a range from 0 to1, as this value increases better clustering results are achieved.

## 5. EXPERIMENT & RESULTS

In this section we discuss our experimental results for evaluating the performance of our proposed clustering approach. The experiments are divided into two parts. The first part works on labeled dataset in order to evaluate the applicability of the proposed system. The second part works on unlabeled dataset in order to measure the scalability of the proposed system and the efficiency of applying it on large datasets.

For labeled dataset we compared our approach against applying bisecting k-means clustering using Hotho method [11], Recupero [4] lexical categories method, lexical nouns methods and stemmed "without ontology" method. For Hotho method we applied the "add" concept method. In this method for each term in the vector representation of the dataset, we check if it appears in the WordNet [5]. In case it appears we extend it by WordNet hypernym synset concept up to 5 levels. Also, we applied first concept of WordNet for word sense disambiguation. To prepare dataset for lexical categories we used WordNet ontology to identify the lexical category of the term. In the following sub-sections we will provide more details about the datasets and the experimental results.

### 5.1 Labeled Dataset

We used Reuters-21578 dataset [24]; it contains 21578 news documents which have been classified into topics. However, not all documents have a topic, so we used only documents that have at least one topic. Figure (4) displays the classes' sizes of Reuters-21578 dataset.

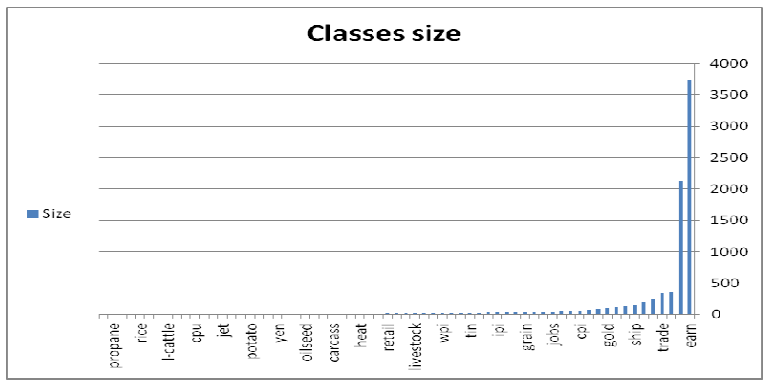

Figure 4 Classes sizes of Reuter dataset

As shown in Figure (4) the dataset has classes of different sizes, some classes have large size such as earn class that has 3734 documents belonging to it, while other classes have size less than 5 documents such as rice. So, we extracted 3 unbiased subsets from the dataset. The first dataset is "Reu-15-20", that represents documents of classes that have size equal 15 and less than or equal 20. The second dataset is "Reu-100-SS" that represents classes which have size greater than or equal 100; we considered only 100 news documents for each class. The third dataset is "Reu-200-SS" contains classes which have size greater than or equal 200; we considered only 200 news document for each class.

### 5.1.1 Labeled Dataset Results

Figure (5) displays F-measure results. Lexical categories representations outperform other methods of dataset representation for datasets Reu-15-20, Reu-200-SS. In case of the dataset Reu-100-SS, lexical nouns representation outperforms other methods of dataset representations. The experimental results of F-measure confirm that Utilizing WordNet ontology for representing document terms either as lexical categories or lexical nouns only outperforms stemmed method "without ontology" and Hotho method. The reason for this result is due to representing semantic relations between document terms; for example, documents that contain words such as rice or wheat will be related to a document with the term "corn" appearing in it; hence the utilization of ontology enhances document clustering results.

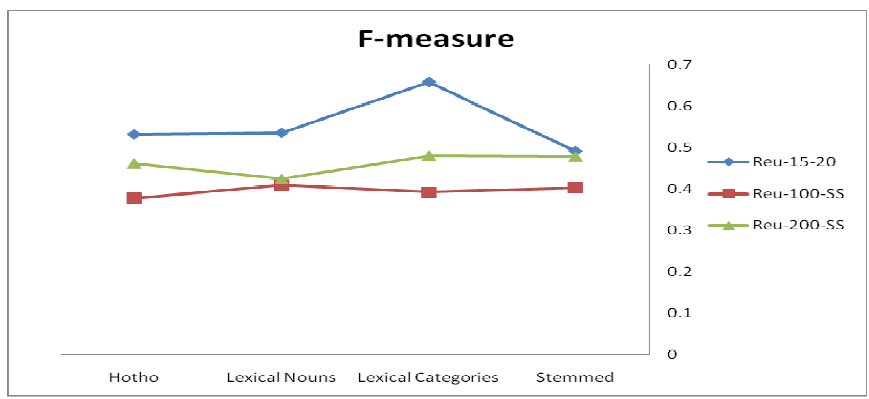

Figure 5 F-measure results of labeled dataset

Figure 6 displays Entropy results, the experimental results for datasets Reu-15-20, Reu-100-SS indicates that, lexical categories representations outperforms other methods of dataset representation. For dataset Reu-200-SS, Hotho representation outperforms other methods of dataset representations.

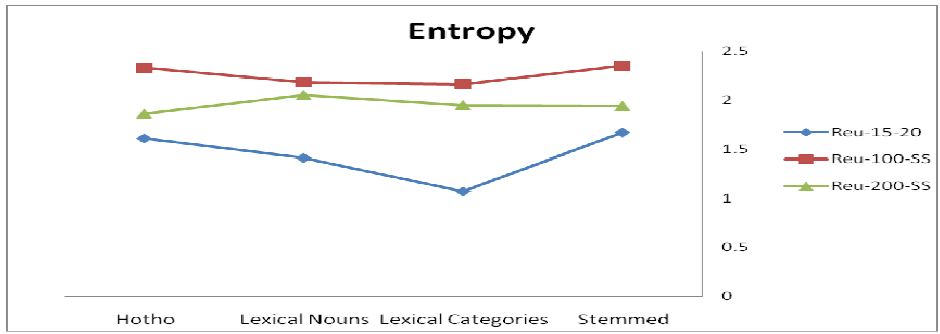

Figure 6 Entropy results of labeled dataset

The experimental results confirm that exploiting ontology for dataset representation enhances document clustering, either by using lexical categories representation or Hotho representation. Figure 7 displays Purity results, for datasets Reu-15-20, Reu-200-SS lexical categories representations outperforms other methods of dataset representation. For dataset Reu-100-SS, Stemmed "without ontology" representation outperforms other methods of dataset representations, but still lexical categories, and lexical nouns representation purity results are near to it.

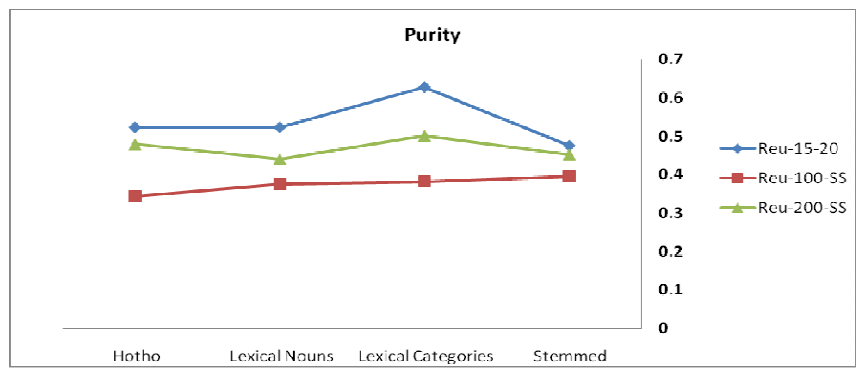

Figure 7 Purity results of labeled dataset

## 5.2 Unlabeled Dataset

We used three bags of words datasets which are available at UCI Machine Learning Repository [18]. The first dataset is NIPS (Neural Information Processing Systems), it consists of 1500 documents, number of words 12419, and number of stems 9072. The second dataset is NYTimes news articles; it consists of 300.000 news articles, number of words 102660, and number of stems 84435. The third dataset is PubMed abstracts dataset which consists of 8200000 records, 141043 words, and 119725 stems.

### 5.2.1 Unlabeled Dataset Results

We run three experiments for NIPS dataset, one for "Without Ontology" method, and the others for the WordNet lexical categories, and WordNet lexical nouns methods. As displayed in Table (1): representing dataset features as WordNet lexical Nouns only results in better internal evaluation than using Stemmed "Without Ontology" or lexical category.

Table 1: Internal Evaluation for Unlabeled Datasets

| Dataset | Type | Features | Internal Evaluation |
|---------|------|----------|---------------------|
| Nips | Without Ontology | 9072 | 0.78092 |
| | WordNet Lexical Categories | 45 | 0.73119 |
| | WordNet Lexical Nouns | 27 | 0.85828 |
| NYTimes | Without Ontology | 84435 | NA |
| | WordNet Lexical Categories | 45 | 0.70215 |
| | WordNet Lexical Nouns | 27 | 0.74354 |
| PubMed | Without Ontology | 119725 | NA |
| | WordNet Lexical Categories | 45 | 0.62839 |
| | WordNet Lexical Nouns | 27 | 0.69913 |

The reasons for this result is due to nouns are more representative for document contents than other parts of speech. Further, using nouns alone provides improved document clustering [13].The results showed that using lexical nouns only get the highest value of internal evaluation.

# 6. CONCLUSION

This paper investigates approaches that enhance and facilitate clustering huge number of documents. Enhancement has been achieved through exploiting WordNet Ontology to represent semantics between documents' terms, in addition to decreasing document features by mapping documents terms to its WordNet lexical categories. On the other hand, this paper introduces an implementation of bisecting k-means algorithm using MapReduce programming model; which facilitates running document clustering in distributed environment. We concluded that using WordNet lexical categories to represent document terms has many benefits, first it represents semantic relations between words, second it reduces features dimensions, and finally it made clustering big data visible by dealing with reduced number of dimensions. Also in this paper we investigated using WordNet lexical categories for nouns only, and we concluded that it enhances the internal evaluation measure. For future work we plan to extend our approach to other lexical categories and to further investigate possible enhancements for bisecting k-means implementation over Map-Reduce paradigm.

## REFERENCES

[1] Zamir, O., & Etzioni, O. (1999). Grouper: a dynamic clustering interface to Web search results. Computer Networks, 31(11), 1361-1374.

[2] Andrews, N. O., & Fox, E. A. (2007). Recent developments in document clustering. Tech. rept. TR-07-35. Department of Computer Science, Virginia Tech.

[3] Cheng, Y. (2008). Ontology-based fuzzy semantic clustering. In Convergence and Hybrid Information Technology, 2008. ICCIT'08. Third International Conference on (Vol. 2, pp. 128-133). IEEE.

[4] Recupero, D. R. (2007). A new unsupervised method for document clustering by using WordNet lexical and conceptual relations. Information Retrieval, 10(6), 563-579.

[5] Miller, G. A. (1995). WordNet: a lexical database for English. Communications of the ACM, 38(11), 39-41.

[6] Dean, J., & Ghemawat, S. (2008). MapReduce: simplified data processing on large clusters. Communications of the ACM, 51(1), 107-113.

[7] Gruber, T. R. (1991). The role of common ontology in achieving sharable, reusable knowledge bases. KR, 91, 601-602.

[8] Cantais, J., Dominguez, D., Gigante, V., Laera, L., & Tamma, V. (2005). An example of food ontology for diabetes control. In Proceedings of the International Semantic Web Conference 2005 workshop on Ontology Patterns for the Semantic Web.

[9]   Ashburner, M., Ball, C. A., Blake, J. A., Botstein, D., Butler, H., Cherry, J. M., ... & Sherlock, G. (2000). Gene Ontology: tool for the unification of biology. Nature genetics, 25(1), 25-29.

[10]  Lauser, B., Sini, M., Liang, A., Keizer, J., & Katz, S. (2006). From AGROVOC to the Agricultural Ontology Service/Concept Server. An OWL model for creating ontologies in the agricultural domain. In Dublin Core Conference Proceedings. Dublin Core DCMI.
      [11]Hotho, A., Staab, S., & Stumme, G. (2003). Ontologies improve text document clustering. In Data Mining, 2003. ICDM 2003. Third IEEE International Conference on (pp. 541-544). IEEE.

[12]  Sedding, J., & Kazakov, D. (2004, August). WordNet-based text document clustering. In Proceedings of the 3rd Workshop on RObust Methods in Analysis of Natural Language Data (pp. 104-113). Association for Computational Linguistics.

[13]  Fodeh, S., Punch, B., & Tan, P. N. (2011). On ontology-driven document clustering using core semantic features. Knowledge and information systems, 28(2), 395-421.

[14]  Cunningham, H., Maynard, D., Bontcheva, K., & Tablan, V. (2002, July). A framework and graphical development environment for robust NLP tools and applications. In ACL (pp. 168-175).

[15]  MacQueen, J. (1967, June). Some methods for classification and analysis of multivariate observations. In Proceedings of the fifth Berkeley symposium on mathematical statistics and probability (Vol. 1, No. 14, pp. 281-297).

[16]  Hartigan, J. A., & Wong, M. A. (1979). Algorithm AS 136: A k-means clustering algorithm. Applied statistics, 100-108.

[17]  Steinbach, M., Karypis, G., & Kumar, V. (2000, August). A comparison of document clustering techniques. In KDD workshop on text mining (Vol. 400, No. 1, pp. 525-526).

[18]  Bache, K. & Lichman, M. (2013). UCI Machine Learning Repository [http://archive.ics.uci.edu/ml]. Irvine, CA: University of California, School of Information and Computer Science.

[19]  Zhao, W., Ma, H., & He, Q. (2009). Parallel k-means clustering based on mapreduce. In Cloud Computing (pp. 674-679). Springer Berlin Heidelberg.

[20]  Huang, A. (2008, April). Similarity measures for text document clustering. In Proceedings of the sixth new zealand computer science research student conference (NZCSRSC2008), Christchurch, New Zealand (pp. 49-56).

[21]  RRNYI, A. (1961). On measures of entropy and information. In Fourth Berkeley Symposium on Mathematical Statistics and Probability (pp. 547-561).

[22]  Deepa, M., Revathy, P. (2012). Validation of Document Clustering based on Purity and Entropy measures. International Journal of Advanced Research in Computer and Communication Engineering, 1(3), 147-152.

[23]  Zhao, Y., & Karypis, G. (2002). Comparison of agglomerative and partitional document clustering algorithms (No. TR-02-014). Minnesota Univ Minneapolis Dept Of Computer Science.

[24]  Lewis, D. D. (1997). Reuters-21578 text categorization test collection, distribution 1.0. http://www. research. att. com/~ lewis/reuters21578. html.

## AUTHORS

**Abdelrahman Elsayed** is working as Research assistant at Central Laboratory for agriculture expert systems. He received his master and BSc in 2008 and 2000 resp. from Information System Dept., Faculty of Computers and Information, Cairo University, Egypt. His research interests are data mining and data intensive algorithms.

**Dr. Hoda Mokhtar** is an Associate Professor in the Information Systems Dept. at the Faculty of Computers and Information, Cairo University. Dr. Hoda received her BSc with "Distinction with Honors" from the Dept. of Computer Engineering, Faculty of Engineering, Cairo University in 1997. In 2000, Dr. Hoda received her MSc degree in Computer Engineering from the Faculty of Engineering, Cairo University. She received her PhD degree in Computer Science from the University of California Santa Barbara (UCSB) in 2005.

**Dr. Osama Ismail** received a MSc. degree in Computer Science and Information from Cairo University in 1997 and a PhD degree in Computer Science from Cairo University in 2008. He is a Lecturer in Faculty of Computers and Information, Cairo University. His fields of interest include, Cloud Computing, Multi-agent Systems, Service Oriented Architectures, and Data Mining.